# Leveraging Digital Twin Technologies: All-Photonics Networks-as-a-Service for Data Center Xchange in the Era of AI [Invited Tutorial]

**HIDEKI NISHIZAWA,[1,*] KAZUYA ANAZAWA,[1] TETSURO INUI,[1] TORU MANO,[1] TAKEO SASAI,[1] GIACOMO BORRACCINI,[2] TATSUYA MATSUMURA,[1] HIROYUKI ISHIHARA,[1] SAE KOJIMA,[1] YOSHIAKI SONE,[1] AND KOICHI TAKASUGI[1]**

[1] *NTT Network Innovation Laboratories, 1-1 Hikarinooka, Yokosuka, 239-0847, Japan*
[2] *NEC Laboratories America, 4 Independence Way, Princeton, New Jersey 08540, USA*
**Hideki.nishizawa@ntt.com*

**Abstract:** This tutorial paper presents a data center exchange (Data Center Xchange, DCX) architecture for all-photonics networks-as-a-service in distributed data center infrastructures, enabling the creation of a virtual large-scale data center by directly interconnecting geographically distributed data centers in metropolitan areas. In contrast to existing vendor-driven optical networking approaches, the proposed architecture adopts an operator-driven and open digital twin paradigm, leveraging cloud-native transponder architectures and open tools/interfaces such as GNPy and CMIS/TAI, and a user-carrier collaborative control framework. In particular, the cloud-native architecture enables operators to flexibly develop, deploy, and manage their own control and automation functions across transponders and controllers using container-based software components. Key requirements for such an architecture in the era of AI are identified: support for low-latency operations, scalability, reliability, and flexibility within a single network architecture; the ability to add new operator-driven automation functionalities based on an open networking approach; and the ability to control and manage remotely deployed transponders connected via access links with unknown physical parameters. We propose a set of technologies that enable digital twin operations for optical networks, including a cloud-native architecture for coherent transceivers, remote transponder control, fast end-to-end optical path provisioning, transceiver-based physical-parameter estimation incorporating digital longitudinal monitoring, and optical line system calibration, demonstrating their feasibility through field validations.

## 1. Introduction

In the 2010s, rapid advancements in high-capacity digital coherent optical networking and virtualization technologies for data center (DC) equipment fundamentally transformed the global communications landscape, enabling cloud-based services such as social networking platforms and video streaming to be delivered on a worldwide scale. In optical transport networks, the commercial adoption of digital coherent transmission around 2010 marked a pivotal shift. Supported by the continuous evolution of semiconductor miniaturization and digital signal processing techniques, coherent transceivers (TRxs) drastically improved both transmission capacity and device compactness. As a result, not only traditional telecommunication carriers but also a diverse range of operators increasingly adopted these technologies. From the perspective of standardization, industry consortia such as the optical interworking forum (OIF), Open ROADM multi-service agreement (MSA), telecom infra project open optical & packet transport (TIP OOPT), and OpenConfig have actively driven the standardization of specifications and control APIs to ensure multivendor interoperability. Meanwhile, from the perspective of network design and operation, research on optical signal propagation design leveraging Gaussian Noise (GN) models has progressed significantly [1, 2], with extensive field-trial validations conducted by various operators [3]. Collectively, these developments have laid a robust foundation for deploying large-capacity optical networks across a wide variety of emerging use cases.

Around the same period, within DCs, the disaggregation of hardware and software—initially applied to servers and later extended to network switches—rapidly advanced and became widely adopted, leading to significant improvements in infrastructure efficiency. The advantages of this approach extend far beyond capital expenditure reduction. By leveraging open hardware and open-source software, cloud operators have been able to rapidly build systems while ensuring both high reliability and scalability. Specifically, new control-plane and service features can be deployed without disrupting ongoing traffic, enabling secure and reliable software rollouts within hours. Cloud-scale telemetry and fully automated failure mitigation enhances service continuity by rapidly detecting and localizing anomalies. A unified software-defined network (SDN) based architecture governs switching platforms, eliminating operational inconsistencies and reducing the likelihood of configuration-induced failures [4].

These technological innovations have accelerated the widespread adoption of cloud services; however, as generative AI and AI-enabled services continue to proliferate, new challenges are emerging. To provide low-latency user experiences, cloud operators increasingly deploy applications from DCs located closer to end users. In addition, as digital twin technologies—which replicate physical objects and environments in the virtual domain using data collected from the real world—become more widely deployed, the need for high-capacity, low-latency networks that efficiently leverage edge DCs located close to end users will continue to expand. For example, in manufacturing environments, various sensor data such as biometric signals, control information, temperature, and $CO_2$ concentration can be continuously synchronized with a digital twin in real time. Such systems enable predictive maintenance functions that help prevent critical incidents before they occur and support remote operations in scenarios where experienced technicians are not physically present on site, demonstrating that digital-twin-driven business use cases have already begun to emerge. In optical networks, technologies are being introduced that allow for monitoring the network across a wider range of characteristics and properties related to optical transmission, thereby improving automated management through telemetry. Furthermore, with respect to network resilience, the intensification of natural disasters driven by climate change and the heightened geopolitical risks are making infrastructure failures harder to predict, while simultaneous failures across multiple locations are becoming more frequent [5]. Consequently, traditional static redundancy alone is no longer sufficient to address such unpredictable disruptions; instead, new approaches—such as dynamic recoverability—are demanded to ensure resilient and reliable service continuity [6].

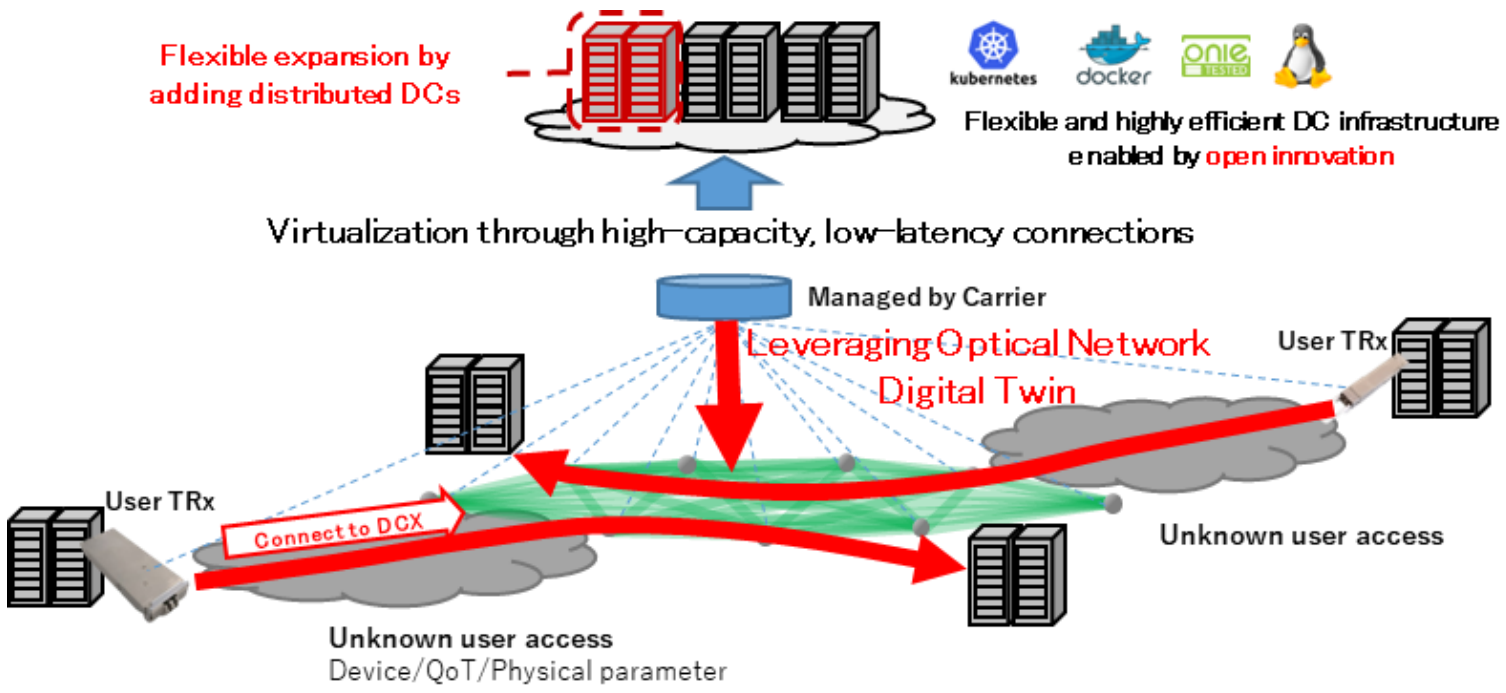

Fig. 1. Data center exchange leveraged by optical network digital twin.

Against the backdrop of two technological innovations— digital coherent optical networking and virtualization—and the environmental changes described in the previous paragraph, expectations are growing for All-Photonics Networks-as-a-Service (APNaaS), realizing a virtual large-scale DC composed of numerous DCs distributed across a metropolitan area, as illustrated in Fig. 1. APNaaS services enable on-demand interconnection of TRxs belonging to a wide variety of users—including carriers, DC operators, and cloud providers—across many-to-many DC sites deployed over an area, by leveraging optical wavelength paths. These TRxs may span heterogeneous form factors, such as CFP modules commonly used by carriers and QSFP-DD modules widely adopted in DCs and cloud environments. By utilizing APNaaS, carriers, DC operators, and cloud providers can virtually operate geographically distributed metro-area DCs as a single large-scale DC. According to [7], the volume of inter-DC traffic in Japan increased by approximately sixfold, from 27 exabytes (EB) in 2013 to 166 EB in 2023, over a ten-year period, and is projected to reach 962 EB by 2033. To date, operators have independently constructed and operated closed DC interconnection (DCI) networks by directly connecting dark fibers, but the proposal of a sustainable infrastructure architecture capable of accommodating future traffic growth has become a need and a challenge. By providing highly multiplexed transmission systems as carrier services using wavelength-division multiplexing (WDM) technology, APNaaS enables sharing of line-system resources among diverse users, significantly improving infrastructure efficiency while allowing scalable expansion in response to user demands.

To address the challenges of APNaaS, this paper presents ongoing initiatives toward an optical network digital twin (ONDT), digitally replicating a physical optical network in a cyber environment to support rigorous optimization of design, operation, and management. For cloud-native architectures of coherent TRx systems, a TRx control architecture based on container technologies has been developed, which is highly compatible with the operational and management frameworks used in data centers [8]. With respect to remote transponder control, mechanisms for remote monitoring and control of user equipment by APNaaS operators have been proposed and validated through field trials, including fiber segments deployed underground and aerial cables [9]. Regarding fast end-to-end (E2E) optical path provisioning, architecture, protocol, and design methods have been proposed to optimize the transmission parameters of TRxs deployed at user sites according to the quality of transmission (QoT), and these methods have been demonstrated over field fibers in densely populated urban areas [10]. For TRx-based physical-parameter estimation technologies incorporating digital longitudinal monitoring (DLM) [11], optical link/network tomography [12, 13], and optical line system (OLS) calibration, vendor-agnostic, open, and automated techniques for line system parameter monitoring/estimation and optimization have been proposed. Their effectiveness and practicality have been experimentally validated in both metro and long-haul field environments [14, 15]. While digital twin concepts have been explored in optical networking, the proposed approach is distinguished by its integration with operator-driven and open architectural principles, as well as user–carrier collaborative control. In addition, it extends digital twin operation beyond transponders to include line systems, thereby enabling unified end-to-end modeling and optimization across the entire optical network infrastructure.

In this paper, we focus on DCX as a use case of APNaaS in distributed DC infrastructures. DCX aims to flexibly and efficiently interconnect geographically distributed DCs using a cloud-native approach [10]. This is achieved by leveraging ONDT technologies, in which physical phenomena are modeled and optimized in the digital domain and the results are fed back into the actual operational environment. The main contributions of this paper are summarized as follows:

- We identify the requirements and challenges for DCX from the perspectives of transponders, controllers, and OLSs. We then present a novel DCX architecture, highlighting its distinguishing characteristics with respect to existing optical networking approaches, including its operator-driven and programmable design, user–carrier collaborative control framework, and support for alien access links.

- We provide a comprehensive overview of the key ONDT technologies required to enable this architecture, and demonstrate how they are integrated into a unified digital twin framework spanning both transponders and line systems. By summarizing field-fiber demonstration results obtained to date, we also clarify the practical feasibility and technological readiness of the proposed approach.

The content of the article is structured as follows. In Sec. 2, we discuss the issues of current network/device architectures and optimization technologies in building infrastructure for generative AI and AI-enabled services. In Sec. 3, we present the corresponding requirements and challenges, along with the architecture we propose. Sec. 4 breaks down the challenges into the transponder, controller, and line system aspects, and describes the enabling technologies and their validation status. Sec. 5 discusses future research directions, and Sec. 6 concludes the paper.

## 2. Issues in existing network/device architectures and optimization technologies

Fig. 2 illustrates the architectures and issues related to currently existing high-capacity optical networks/devices. In Fig. 2(a), the hierarchical ring, enabled by reconfigurable optical add-drop multiplexer (ROADM) deployment, offers excellent scalability and ensures high reliability by applying protection mechanisms that rapidly switch paths upon failures while leveraging ring topologies. Therefore, this architecture has been widely used as carrier networks optimized for efficiently and reliably multiplexing consumer services such as Fiber-to-the-Home (FTTH) and mobile. However, due to the hierarchical structure, fiber routes tend to become longer (e.g., the orange line in the figure represents an inter-prefecture connection), and packet buffering in routers—introduced to efficiently aggregate a large number of consumer connections—can easily increase latency. As a result, it is difficult to satisfy the low-latency requirements that are critical for inter–DC communication. In Fig. 2(b), Regional DCI directly interconnects DCs with dark fibers, which enables low latency and provides strong reliability through full-mesh DC-to-DC connectivity. On the other hand, its scalability is limited because each time a new DC is added, fiber connections must be provisioned from that site to all existing locations [16]. Moreover, this architecture assumes scenarios where fiber resources are already abundant or where new fiber deployment is relatively easy. In Fig. 2(c), The routed hub-and-spoke architecture has the advantages of operational simplicity—being directly applicable to L2/L3 network topologies—and reduced initial deployment cost. However, power consumption and redundancy-related risks under failure conditions are concentrated at the central hub site, and low spectral efficiency of the fiber infrastructure prevents scalable capacity expansion. Although fiber propagation delay is relatively small compared to that in Fig. 2(a), if the router throughput at the hub is insufficient, buffering-induced latency can still become significant. Next, we discuss device architecture. Fig. 2(d) illustrates the issues of optical network devices. While DC devices have increasingly adopted open platforms enabling autonomous operation and digital-twin-based management, optical network devices have traditionally been developed based on vendor-specific architectures under vertically integrated business models. As a result, both control interfaces and operational tools are generally proprietary to each vendor. Consequently, there remain issues in terms of flexibility and scalability when operators attempt to apply automatic and digital-twin-based operation to optical network devices, including those already implemented.

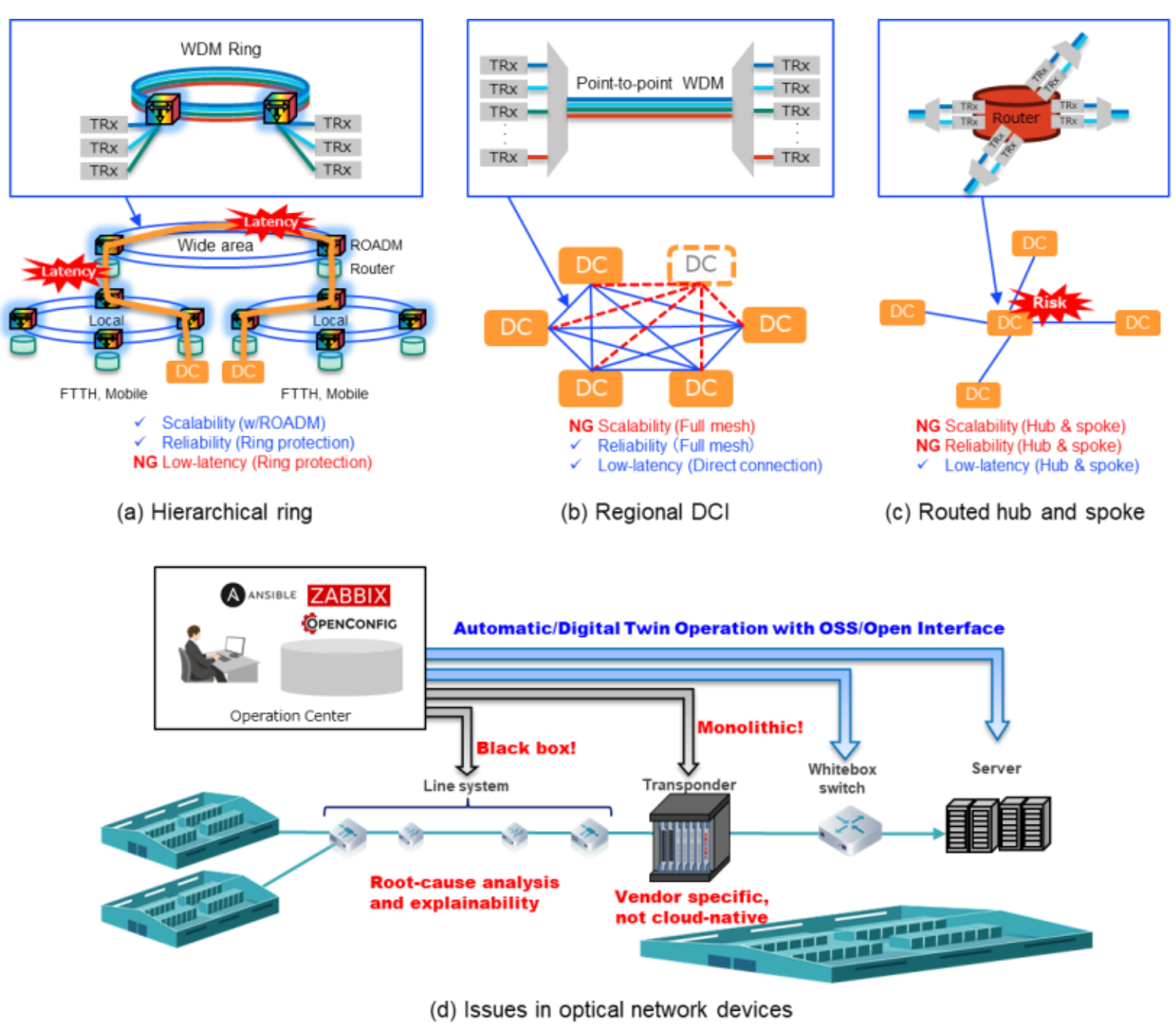

Fig. 2. Existing large-capacity optical network/device architectures and these issues.

Finally, we discuss the issues of optimization technologies. To efficiently accommodate rapidly increasing traffic, new network optimization technologies have been actively studied. Optical Spectrum as a Service (OSaaS) is recognized as a model for sharing optical spectrum and fiber infrastructure among multiple operators [17]. OSaaS allows customers (service providers) to deploy and operate their own TRxs while dynamically allocating spectral resources over existing infrastructure. In [18], concatenated generalized signal-to-noise ratio (GSNR) profile approach is proposed to estimate the E2E GSNR by measuring the GSNR of each link that constitutes the E2E route with probe channels. However, in [18], GSNR is estimated for each wavelength by using a probe channel at the same wavelength as the operational signal light, which results in significant provisioning time in the case of multi-wavelength multiplexed transmission. Moreover, OSaaS has only been studied for P2P connections, and optimization methods for complex topologies—such as those interconnecting distributed data centers—have not been investigated. Regarding autonomous operation and the application of digital twin technologies, [19, 20] reports field-trial demonstrations of autonomous optical networking (AON) enabled by the combination of large language models (LLMs), digital twins, and autonomous control, achieving zero-touch lifecycle management. Such autonomous capabilities can potentially enable flexible interconnection among distributed sites while improving scalability and reliability using advanced technologies such as machine learning (ML). However, realizing AON requires the integrated deployment of numerous components including sensors and actuators. In addition, the application of black-box approaches such as ML-based decision-making introduces deviations from conventional operational models, raising concerns about root-cause analysis and explainability, especially in the event of failures [15, 21].

In this paper, we focus on coherent transmission systems. While Intensity Modulation-Direct Detection (IM-DD) based technologies are widely used for short- and mid-reach interconnections, their design principles and performance modeling differ fundamentally from those of coherent systems. Therefore, to maintain clarity and focus, this tutorial is limited to coherent amplified optical networks. Extending the framework to IM-DD systems is an interesting direction for future work.

## 3. Architecture, requirements, and challenges for DCX

In this section, we describe the proposed architecture, along with the main requirements and challenges necessary to address the problems illustrated on Sec. 2.

From the standpoint of technology dissemination, broader adoption and network scalability, we consider the active employment of open and disaggregated approaches. Considering the requirements that will be defined in detail later, the presented study considers methods based on an analytical physical model. That said, this choice does not exclude ML-based methods; in fact, the effective integration of ML remains an important prospect for future work.

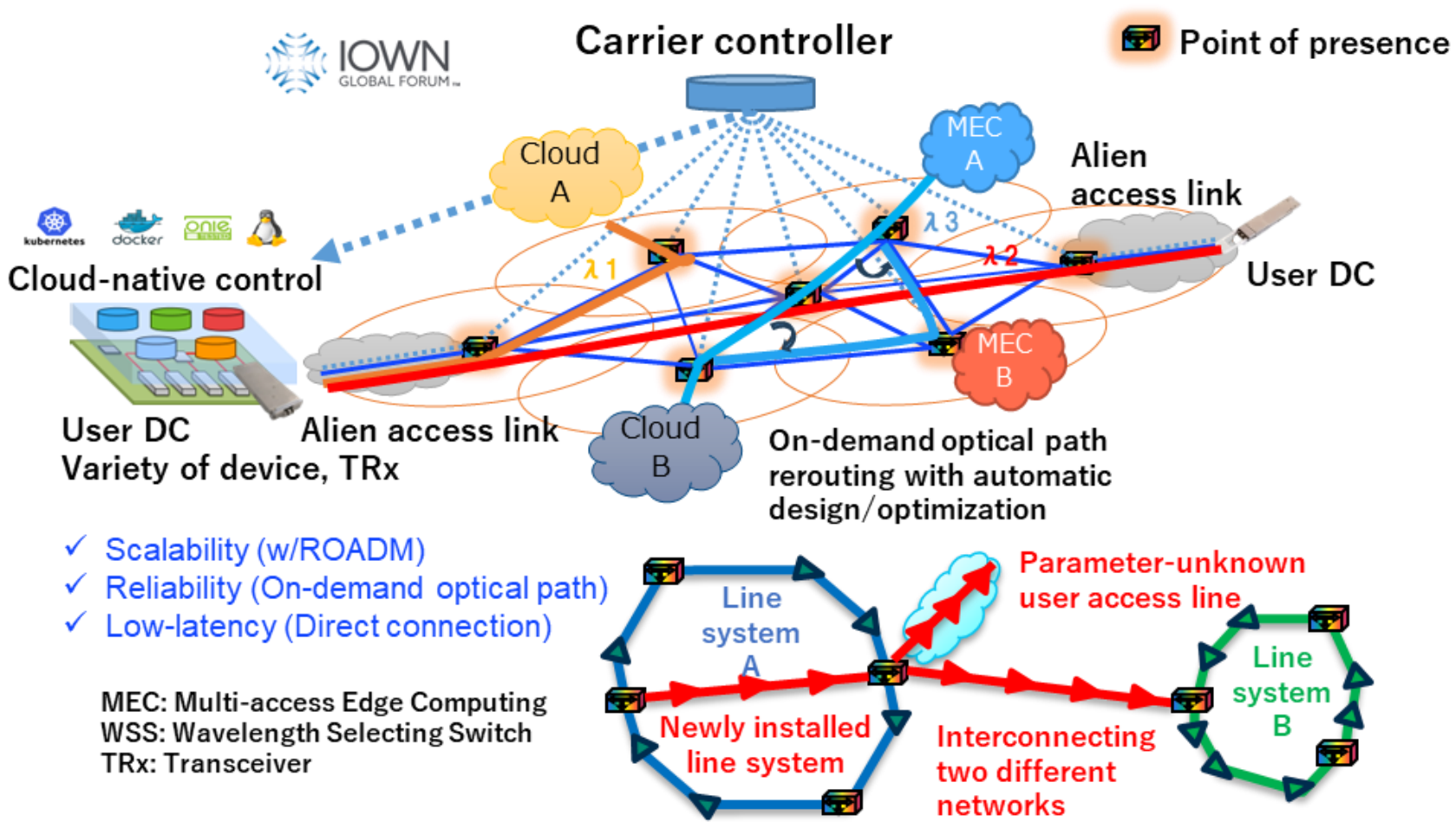

Fig. 3. Proposed DCX architecture.

Fig. 3 shows our proposed DCX architecture. A point of presence (POP) equipped with ROADM technology can add or drop arbitrary wavelengths from a wavelength-multiplexed optical signal by combining a demultiplexer, a wavelength-selective switch (WSS), and a multiplexer. Unlike full-mesh or hub-and-spoke architectures, the proposed DCX connects POPs directly via fiber links and centrally controls them with a carrier controller (The role sharing between users and carrier will be discussed later). As illustrated by the examples of lambda 1~3, optical signals from many sites can be efficiently aggregated using WDM, thereby ensuring high network scalability. Moreover, in contrast to hierarchical ring networks, DCX provides low-latency service delivery by establishing direct optical WDM paths between sites. As shown for lambda 3 in Fig. 3, the introduction of automated optical-path design and operation technologies further enables on-demand rerouting, offering the potential to achieve high reliability. Subsequently, by realizing a device architecture that permits cloud-native control from the carrier controller, we facilitate APNaaS-based connectivity among arbitrary users. Prospectively, our objective is to allow APNs to be configured instantaneously in accordance with machine requirements—such as GPUs, memory, and other resources within the user data center.

To clarify the architectural positioning of DCX, we highlight its key design principles in contrast to existing multi-vendor optical networking solutions. While recent approaches have improved interoperability, many existing systems remain largely vendor-driven, with control and optimization functions embedded within proprietary implementations. In contrast, the proposed DCX architecture adopts an operator-driven and open digital twin paradigm, in which control, optimization, and automation functions are externalized from network devices and implemented at the operator level. In addition, DCX introduces a user–carrier collaborative control framework, where end users and network operators jointly participate in optical path provisioning through standardized information exchange (e.g., TRx characteristics and QoT-related parameters). This enables coordinated decision-making across administrative domains, which is particularly important in scenarios involving user-deployed transponders connected via alien access links with unknown parameters. Furthermore, DCX is built upon an open ecosystem leveraging open tools and interfaces such as GNPy for physical-layer modeling, Common Management Interface Specification (CMIS) and Transponder Abstraction Interface (TAI) for TRx control. In addition, coherent transponders are designed with a cloud-native architecture, where control and monitoring functions are implemented as containerized software components. This enables flexible deployment of operator-defined functionalities, seamless integration with data center operational frameworks, and rapid extensibility of automation features.

In the following, we analyze the challenges of DCX. Although it offers low-latency operations, scalability, reliability, and flexibility with a single network architecture, its implementation requires technologies that can rapidly interconnect and optimize a wide variety of devices—such as TRxs and line systems—owned by end users, cloud service providers, and multi-access edge computing (MEC) operators. With respect to the data plane of coherent TRxs, specifications such as Open ROADM, 400ZR, and OpenZR+ have already been standardized, and interoperable commercial products are widely available. Moreover, the standardization process has recently pushed towards higher rates, such as 800G. However, technologies for the control plane—namely, the control and monitoring of geographically distributed and heterogeneous TRxs—as well as methods for optimizing operating modes across TRxs, have yet to be fully established. Regarding line systems, technologies that enable interoperability among devices from different vendors and rapid provisioning remain immature, making it difficult to optimize optical wavelength paths across multiple line systems as described at the lower of Fig. 3. It is also important to note that both hierarchical ring and regional DCI architectures shown in Fig. 2 (a) and (b) have generally assumed that a single operator installs and operates optical TRxs and line systems within its own infrastructure. In contrast, these devices in DCX are typically deployed at user DCs, outside the carrier building. Therefore, it is necessary to develop technologies and operational methodologies that can support diverse topologies between arbitrary user sites and enable E2E connectivity over alien access links (AALs), where quality and line parameters are unknown. Moreover, directly interconnecting user DCs over optical paths while minimizing electrical conversions is equivalent to reducing the number of parameter monitoring points in the electrical domain. This necessitates the development of new technologies and operational approaches for line system optimization and transmission-quality monitoring. Furthermore, devices deployed at user DCs must adopt an architecture that allows future implementation of new functionalities such as operator-driven automatic/digital-twin operation, and deploy mechanisms to avoid irregular connections caused by accidents, software failures, or malicious attacks, to ensure network security and robustness.

## 4. Challenges and technologies for transponder, controller, and line system

While many of the underlying monitoring and control capabilities exist in modern TRxs and OLSs, the novelty of the proposed approach lies in their integration into an operator-driven digital twin framework. This shift from device-centric to system-level operation is a key enabler of DCX. By combining device-level telemetry with system-level coordination and open interfaces, the proposed architecture enables cross-domain and end-to-end visibility, as well as closed-loop optimization, beyond conventional device-centric operation. In this section, we decompose the requirements of DCX presented in Sec. 3 into challenges related to transponders, controllers, and line systems. Then, we discuss the enabling technologies, supported by results from field experiments.

### *4.1 Transponder*

The key features are cloud-native architecture and remote controllability. A control architecture for TRx must be designed to provide high compatibility with both the OSS systems used in DCs and the equipment installed at the customer premises equipment (CPE) [22, 23]. The challenges and requirements can be summarized in the following three categories.

1) <u>Operation of transponders equipped with different hardware components:</u> A single architecture that efficiently accommodates both carrier-based components and DC cloud-based components with minimum impact on hardware and software.
2) <u>Adaptation to various network operating systems (NOS) and Apps, enabling operator-driven future technology additions and scaling:</u> Hardware and software are disaggregated, agile monitoring and control of hardware components with OSS, facilitating the application of new technologies such as automated control and digital twins.
3) <u>Operation of transponders at user sites:</u> DCX operators can remotely monitor and control user equipment, efficiently performing updates, fault isolation, and preventing irregular connections.

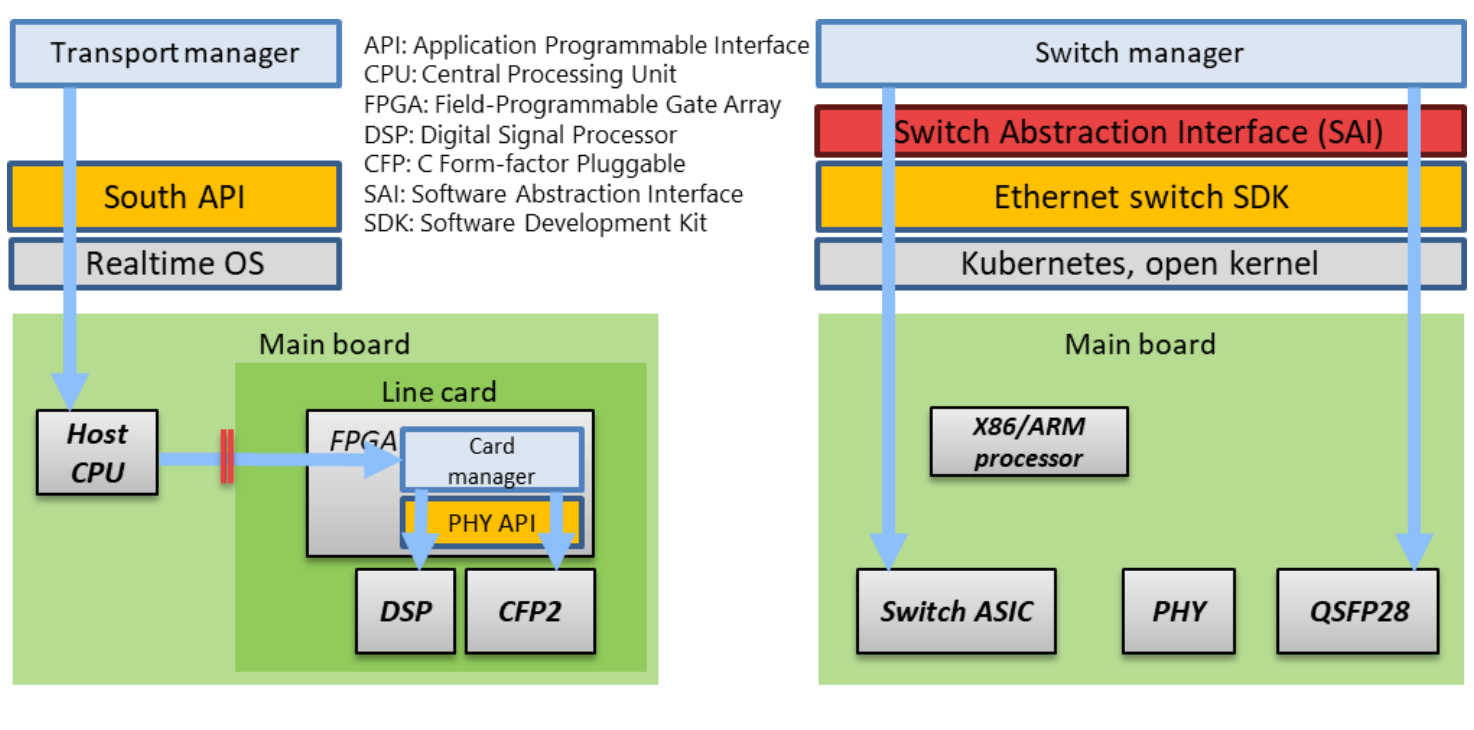

Fig. 4. Comparison of device architecture. (a) Traditional vertically integrated transponder. (b) Whitebox switch decoupled software from hardware.

Traditionally, transponders have been developed using closed, tightly coupled hardware architectures under vertically integrated business models, as they have been required to provide extremely high reliability as the lowest layer of carrier networks. As a result, they have been developed based on proprietary device architectures. Fig. 4 (a) presents the architecture of vertically integrated transponder that typically consists of a main board equipped with a host central processing unit (CPU) and a line card incorporating a field-programmable gate array (FPGA). Distinct application programming interfaces (APIs) are defined for the host CPU and the FPGA, and vendor-specific interfaces are also employed between the main board and the line card. Furthermore, management software is often separated into components to host CPU control and line card control, which tends to make the hardware control logic a black box. From an operator's perspective, this opacity poses a barrier to open innovation and limits the adoption of advanced technologies such as automation and digital twins. In contrast, as presented in Fig. 4(b), the architecture of whitebox switches has embraced a clear separation between hardware and software. In such architectures, components such as switch ASICs and Ethernet interfaces are implemented on a main board equipped with x86 or ARM processors, directly controlled by a unified switch manager through an open kernel, a switch software development kit (SDK), and the Switch Abstraction Interface (SAI) [24].

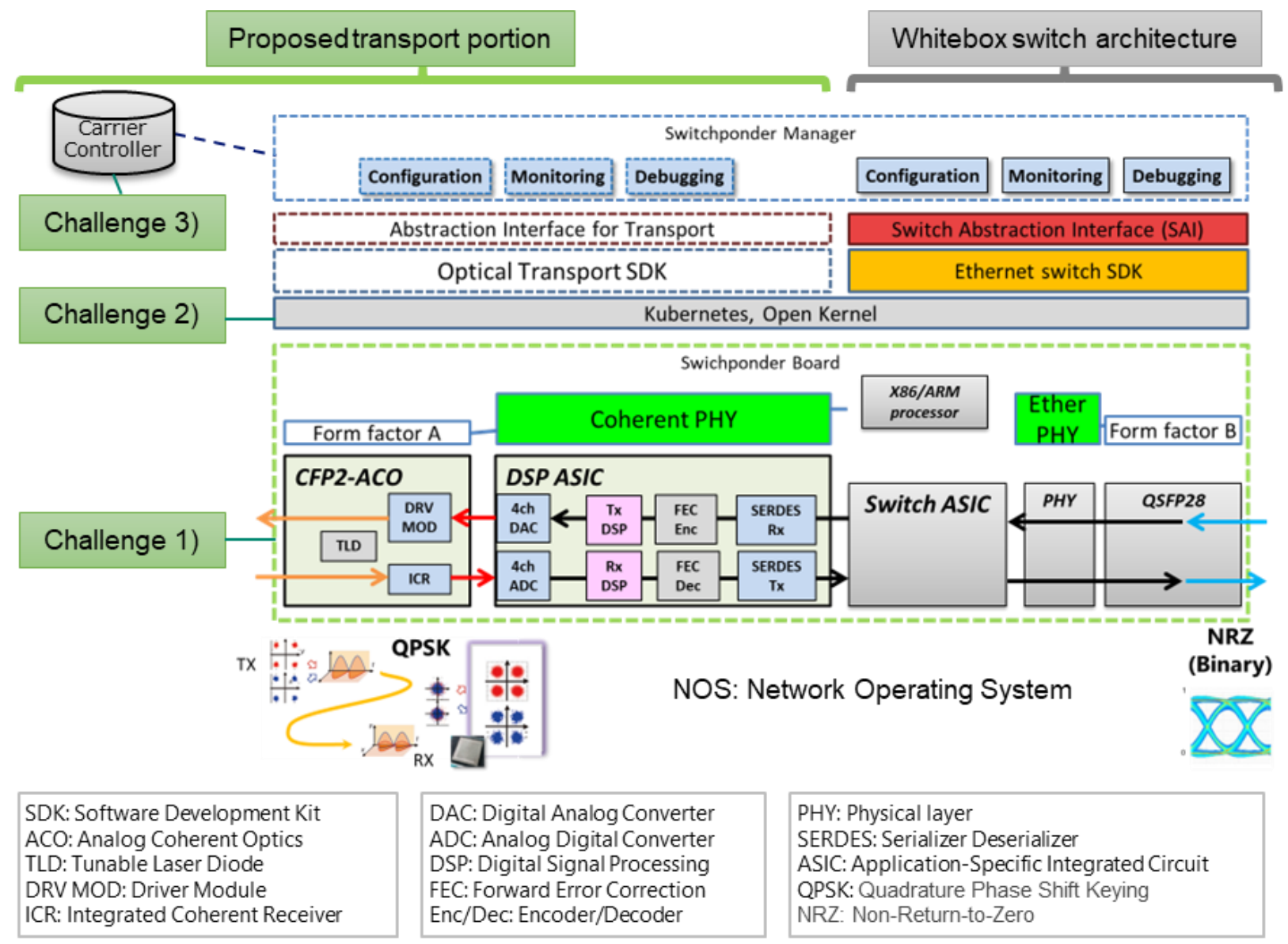

Fig. 5. Proposed cloud-native architecture for coherent TRx. The right-hand side depicts a conventional whitebox switch, in which components are mounted on the board together with an x86/ARM processor for running the open kernel and manager software. The upper-layer manager functions are implemented as user-space libraries of the kernel and directly control these components via a software development kit (SDK). The left-hand side shows the architecture in which a coherent module consisting of a CFP2-ACO and a DSP ASIC is incorporated into the conventional whitebox switch architecture. The CFP2-ACO integrates coherent optical components for transmission and reception. The DSP ASIC is connected to the CFP2-ACO via high-speed analog signals.

Fig. 5 illustrates our proposed architecture in 2019 for a whitebox packet transponder supporting the world's first multivendor/generation TRx capability, developed in collaboration with Edgecore Networks and NTT Innovative Devices [25]. By positioning the CFP2-ACO as an internal form factor of the device in the same manner as QSFP28 modules, and by treating the digital signa processing (DSP) application-specific integrated circuit (ASIC) as a coherent physical layer (PHY), the proposed architecture allows the existing whitebox switch architecture and system specifications to be reused without modification. Although omitted from the figure for simplicity, this architecture also enables the reuse of Open Compute Project (OCP)-compliant open hardware components—such as fans, power supplies, and indicators—that have been widely adopted in whitebox switches.

First, we discuss Challenge 1). The hardware used for coherent communication is highly diverse, depending on factors such as the number of dies integrated into the DSP package, the types of control interfaces, tributary control of optical paths associated with polarization multiplexing and higher-order modulation formats, and the implementation approach of the DSP package or module. For example, with respect to control interfaces, CFP modules, which are widely adopted by carriers are controlled via the Management Data Input/Output (MDIO) interface, whereas control of on-board large-scale integrated circuits (LSIs), such as DSP ASICs, is often performed using the Inter-Integrated Circuit ($I^2C$) interface. The same applies to QSFP-DD modules, which cloud operators widely adopt. Today, for CFP modules, the TAI [8]—defined within TIP to enable hardware–software disaggregation of transponders—has been implemented in the 400-Gbit/s disaggregated muxponder "Phoenix", which has been specified under TIP OOPT [26]. For QSFP-DD modules, standardization efforts are progressing under MANTRA within TIP OOPT [27], based on CMIS and coherent-CMIS (C-CMIS) specifications defined by the OIF. These specifications are being implemented for disaggregated cell site gateways [28].

Next, we discuss Challenge 2). By incorporating open-source software (OSS) technologies into optical transport networks, advanced automation capabilities—similar to those applied to servers and whitebox switches—can be enabled. Furthermore, by combining OSS technologies with hardware abstraction layers such as the TAI, hardware-independent software can be developed, thereby reducing development time and cost dramatically. Based on this concept, TIP OOPT has been promoting the development of Goldstone, an open-source NOS [29]. Goldstone implements widely adopted OSS components across most of its software stack and reduces reliance on proprietary implementations, thereby improving compatibility with software ecosystems commonly used in DC infrastructures and cloud environments.

In [30], experimental results are reported in which ONDT technologies are applied to operate devices equipped with three generations of DSPs, each employing different module implementation approaches, by leveraging TAI and Goldstone solutions. Furthermore, [14] reports experimental results in which the same architecture is applied to a large-scale disaster recovery scenario. In this scenario, operators A and B rapidly restore network and computing service resources by extending virtualization concepts, commonly used for servers and routers, to transponders. When delegating a subset of transponder line ports to other operators, the associated resources are isolated, and access is restricted to only the delegated portion within the NOS. In the reported experiment, resource isolation and access control are implemented by leveraging Goldstone's containerization capabilities. The results presented above indicate that the cloud-native architecture for coherent TRxs successfully addresses Challenges 1) ~ 2).

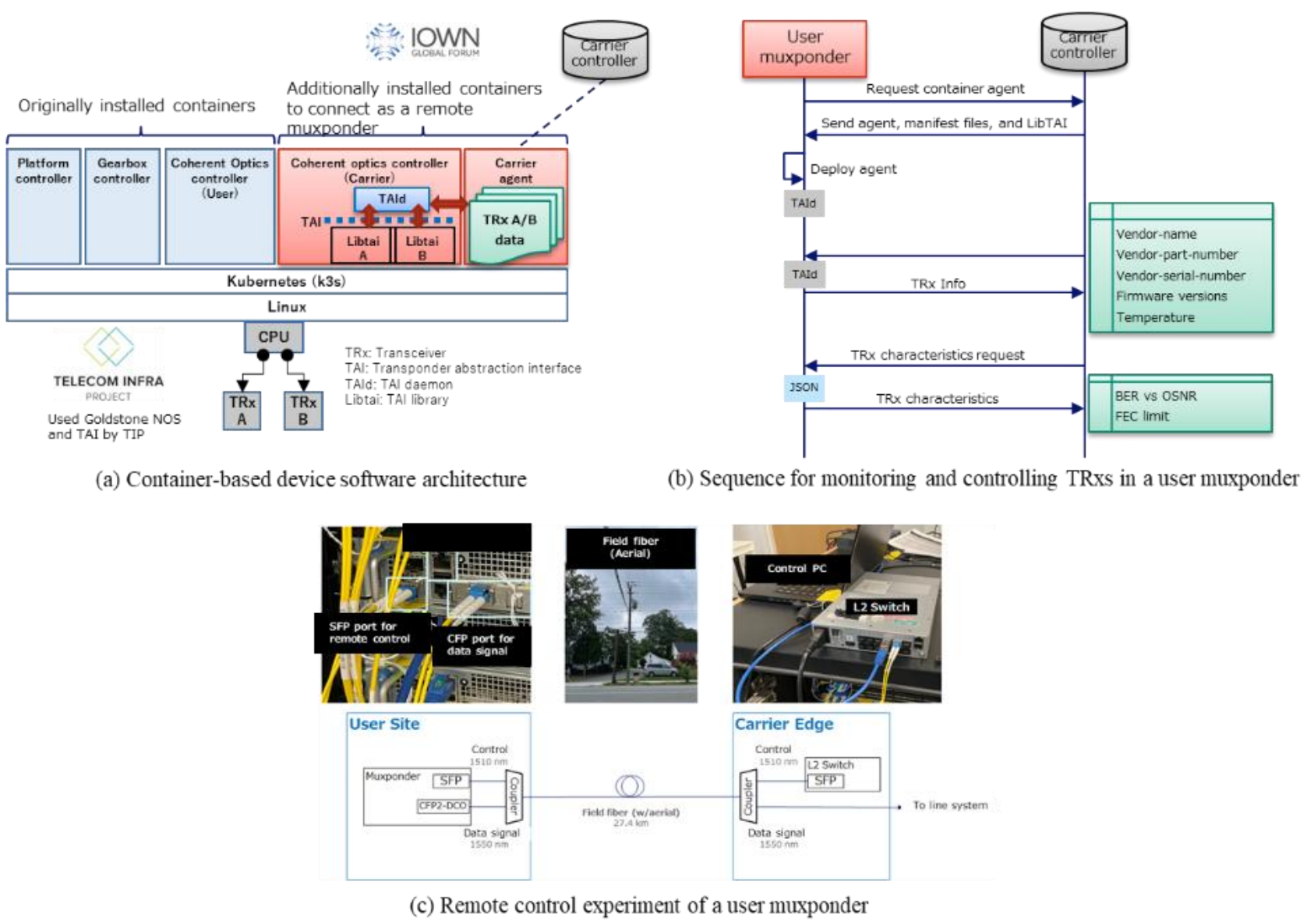

(a) Container-based device software architecture

(b) Sequence for monitoring and controlling TRxs in a user muxponder

(c) Remote control experiment of a user muxponder

Fig. 6. Remotely monitor and control user equipment. (a) shows an example of a container-based device software architecture, and (b) shows an example sequence for monitoring and controlling TRxs in a user muxponder. (c) presents a remote control experiment of a user muxponder. A 27.4-km field fiber link at Duke University's testbed in the United States including a 2.29-km aerial segment was utilized.

Finally, we discuss Challenge 3). When monitoring and controlling a wide variety of TRxs under vertically integrated architectures, mismatches between the specifications of the NOS and the TRxs can lead to operational incidents. In addition, such architectures often prevent operators from leveraging vendor-specific fault diagnosis and troubleshooting functions. Fig. 6(a) illustrates TRx monitoring and control within a user muxponder based on a cloud-native architecture [10]. In this experiment, in addition to the container set initially installed on the user muxponder (shown in blue), a carrier coherent optics controller and a carrier agent (shown in red) were subsequently deployed as containers. This setup assumes a use case in which the carrier performs control of the user muxponder [31]. Fig. 6(b) illustrates an example sequence in which a carrier remotely controls a user muxponder. When the user muxponder is connected to DCX, it retrieves and installs the required containers from the carrier controller, and then transmits TRx-related information to the carrier controller. The carrier controller subsequently verifies whether the TRx is authenticated, and, if so, requests additional information regarding the characteristics of the TRx from the user muxponder. The sequence shown in Fig. 6(b) was designed, implemented, and tested, and the entire sequence—including container startup—was completed within 90 seconds [10]. Through this mechanism, the carrier can apply its own TRx control functions to remotely located user TRxs by deploying them as containers. By managing these functions under its own control, the carrier can prevent irregular connections and thereby ensure network security and robustness. In addition, an experiment was conducted in which the user muxponder was controlled using only the data-signal fiber from the carrier edge, without relying on a dedicated control fiber [9]. Fig. 6(c) shows the remote control experiment of the user muxponder. The data signal (1550 nm) from a CFP2-DCO inserted into the muxponder was wavelength-multiplexed with a Gigabit Ethernet control signal (1510 nm) from a small form-factor pluggable (SFP) module also inserted into the muxponder, and transmitted over the same fiber. Using the sequence in Fig. 6(b), successful control was confirmed, demonstrating that the proposed approach enables remote control using existing equipment. Through these experiments, it was demonstrated that DCX operators can remotely monitor and control user equipment at an advanced level, enabling operations such as software updates and fault isolation. These results confirm that the proposed mechanisms for remote monitoring and control of user equipment satisfies the requirements of Challenge 3).

### *4.2 Controller*

This section proposes the implementation of a technology for the remote design and optimization of the parameters of coherent transceivers installed at user locations. The challenges and requirements can be summarized into the following three categories.

1) <u>Provisioning and rerouting based on user TRx characteristics:</u> A mechanism enabling users and DCX operators to collaborate, allowing carriers to reference user TRx characteristics to achieve optical WDM path design.
2) <u>Supports diverse topologies among arbitrary multi-points:</u> Enabling efficient E2E optical path design in a short time while minimizing errors, remotely calculating optical signal quality degradation and configuring it in the devices.
3) <u>Automatic selection of the optimal mode based on distance and QoT</u>: Enable automatic optimal mode selection across multiple vendors and coherent modes to facilitate user-to-user connections at any distance and fiber quality.

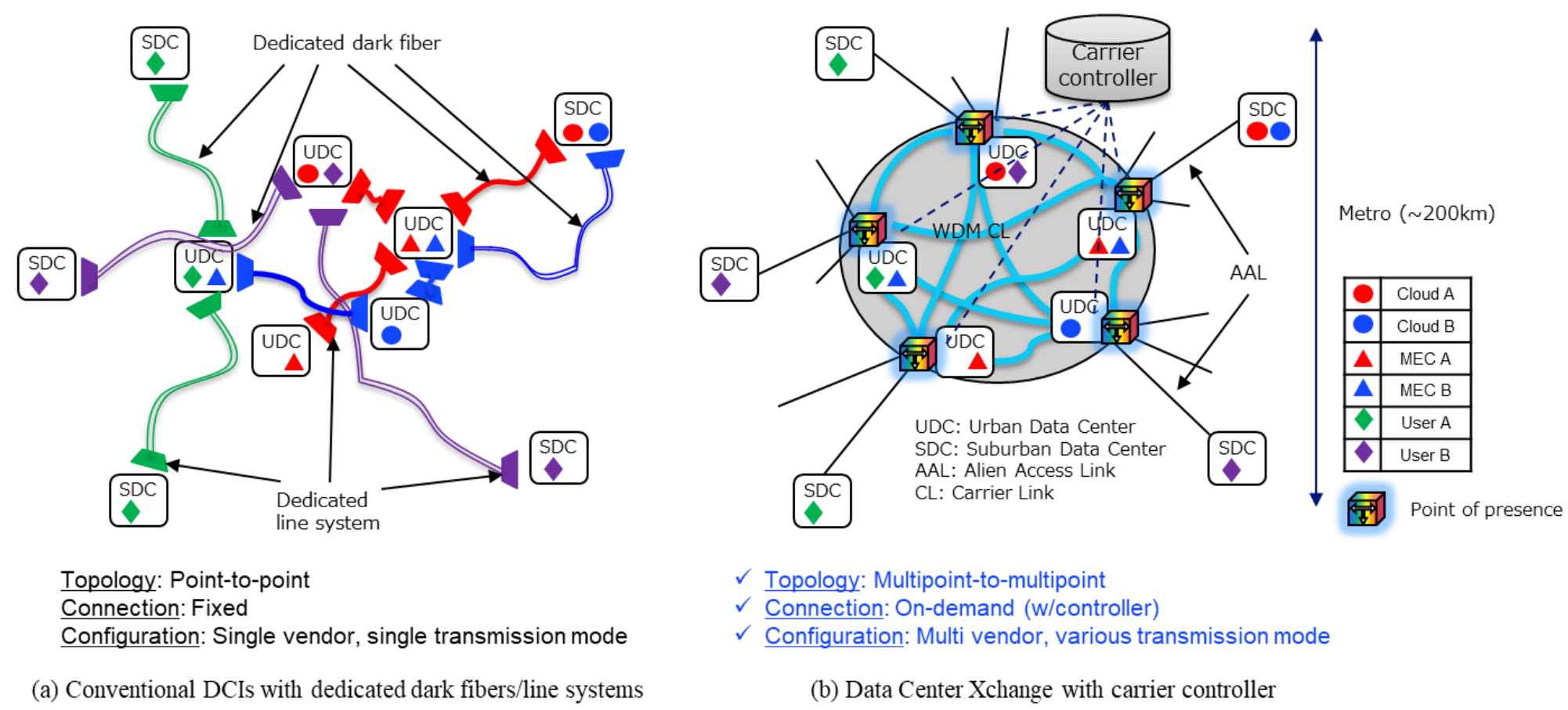

(a) Conventional DCIs with dedicated dark fibers/line systems

(b) Data Center Xchange with carrier controller

Fig. 7. A configuration example and advantage of DCX. (a) presents conventional DCIs. As each operator deploys its own dedicated dark fibers/line systems, network resource utilization is generally inefficient. (b) presents DCX. Under the management of a carrier controller, users can establish on-demand connections between arbitrary DCs by connecting via alien access links to POPs, with high network utilization efficiency enabled by WDM.

Fig. 7 compares conventional DCI with DCX and highlights the advantages of DCX. The Urban DCs (UDCs) are surrounded by multiple Suburban DCs (SDCs). Fig. 7(a) illustrates a conventional scenario in which multiple operators, shown in different colors, independently construct their own DCIs using dedicated dark fibers/line systems. Because the network infrastructure is not shared, overall network utilization efficiency is generally low. Fig. 7(b) shows an example of DCX configuration. The carrier deploys points of presence (POPs) at each UDC, which are interconnected via WDM-based carrier links (CLs) owned by a telecom carrier. User equipment connects from SDCs to UDC POPs via AALs, enabling direct optical connectivity to remote clouds/MECs or other user equipment without electrical conversion. Here, AALs are assumed to be access links for which neither the user nor the carrier knows the link parameters, regardless of ownership (Methods for automated link-parameter extraction are discussed later). In addition, as described in Sec. 4-1, secure channels are pre-established between the user and the carrier, through which control information can be exchanged. With the integration of technologies corresponding to Challenges 1)~3), the DCX architecture supports on-demand multipoint connectivity across heterogeneous TRxs and transmission modes.

First, we discuss the controller architecture and protocols to address Challenge 1). We begin by considering an approach in which end users take the primary role in performing E2E QoT estimation and designing optical WDM paths. In this case, carrier involvement is limited to preparing idle channels for probing and reserving transmission bandwidth on the line system in response to user requests; all other tasks related to optical path establishment are performed solely by the users. While this user-driven approach simplifies the overall system by allowing users to directly implement optical WDM paths, it also has notable limitations. Users are unable to benefit from advanced services such as route switching in the event of failures in the line-system segment, and carriers cannot prevent irregular connections caused by accidents, software failures, or malicious attacks. From this discussion, it becomes clear that a user–carrier collaborative approach to optical WDM path design is essential to satisfy the requirements of DCX. The procedure for optical path establishment is summarized as follows [32]. i) The transmitting user requests the carrier to establish an optical path. ii) The user sends its TRx characteristics to the carrier. iii) The user and carrier synchronously estimate QoT of AAL. iv) The user or carrier selects the TRx mode using the estimated link parameters. v) The user configures the TRx with the selected mode.

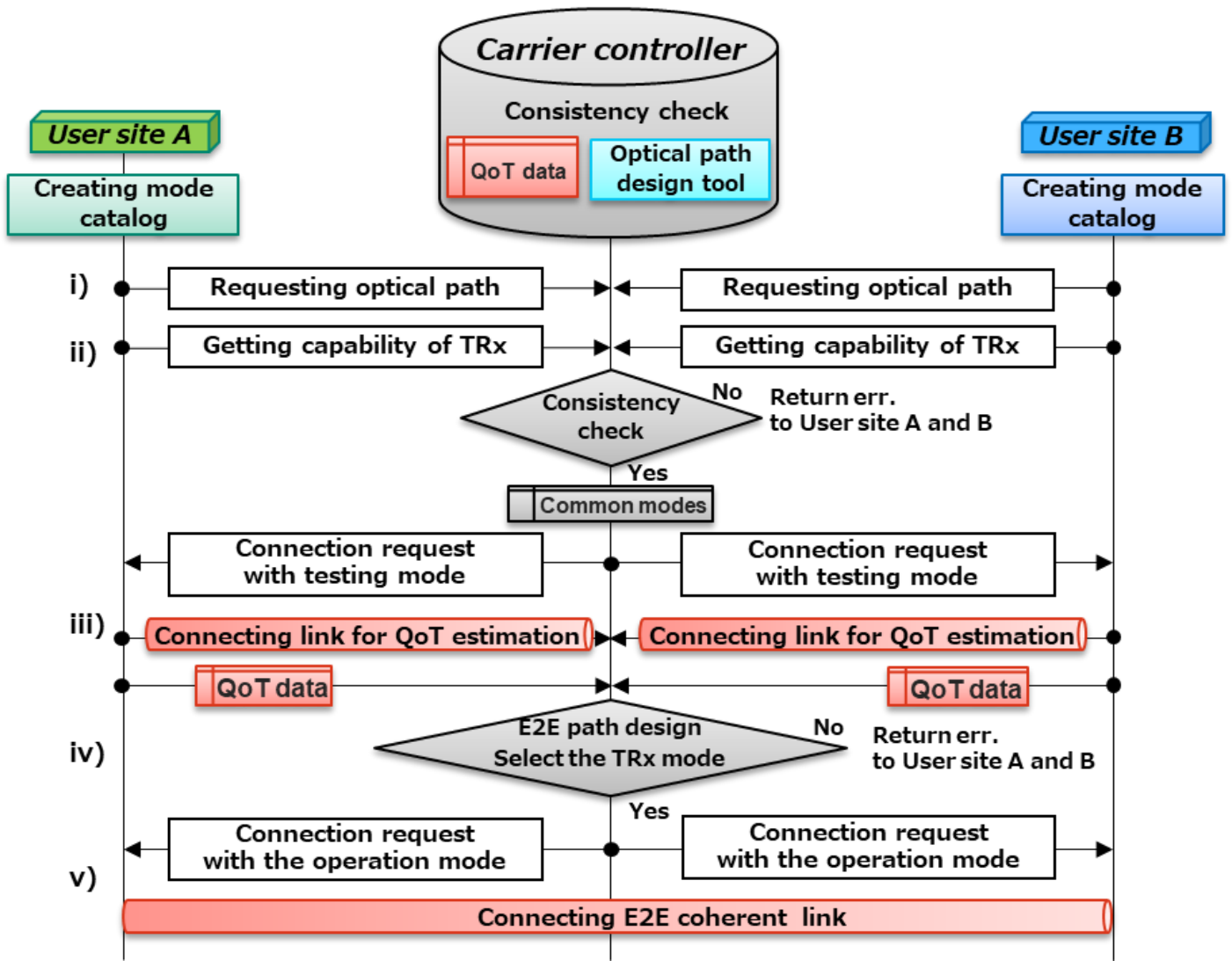

Fig. 8. Proposed architecture and protocol for a user-carrier collaborative approach.

This approach represents a shift from existing single-domain or vendor-controlled provisioning models toward a collaborative control paradigm across user and carrier domains. By explicitly defining the roles of users and operators and enabling standardized information exchange, the proposed framework supports scalable and flexible optical path provisioning in heterogeneous and multi-domain environments. Fig. 8 illustrates our proposed architecture and protocol. Each user site generates a mode catalog containing parameters such as bitrate, modulation format, and FEC type. The carrier controller provides E2E optical path design capabilities, including consistency checks of TRxs located at user sites, a database for storing QoT data, and optical path design tools such as GNPy [33]. The controller compares the mode catalogs received from the TRxs and generates a list of common modes. If no interoperable mode exists, an error is returned to the TRxs. Based on procedures i) ~ v), the user and the carrier collaboratively execute the workflow for E2E optical path provisioning. Thus, for Challenge 1), we have demonstrated a baseline architecture and protocols for user–carrier collaborative provisioning.

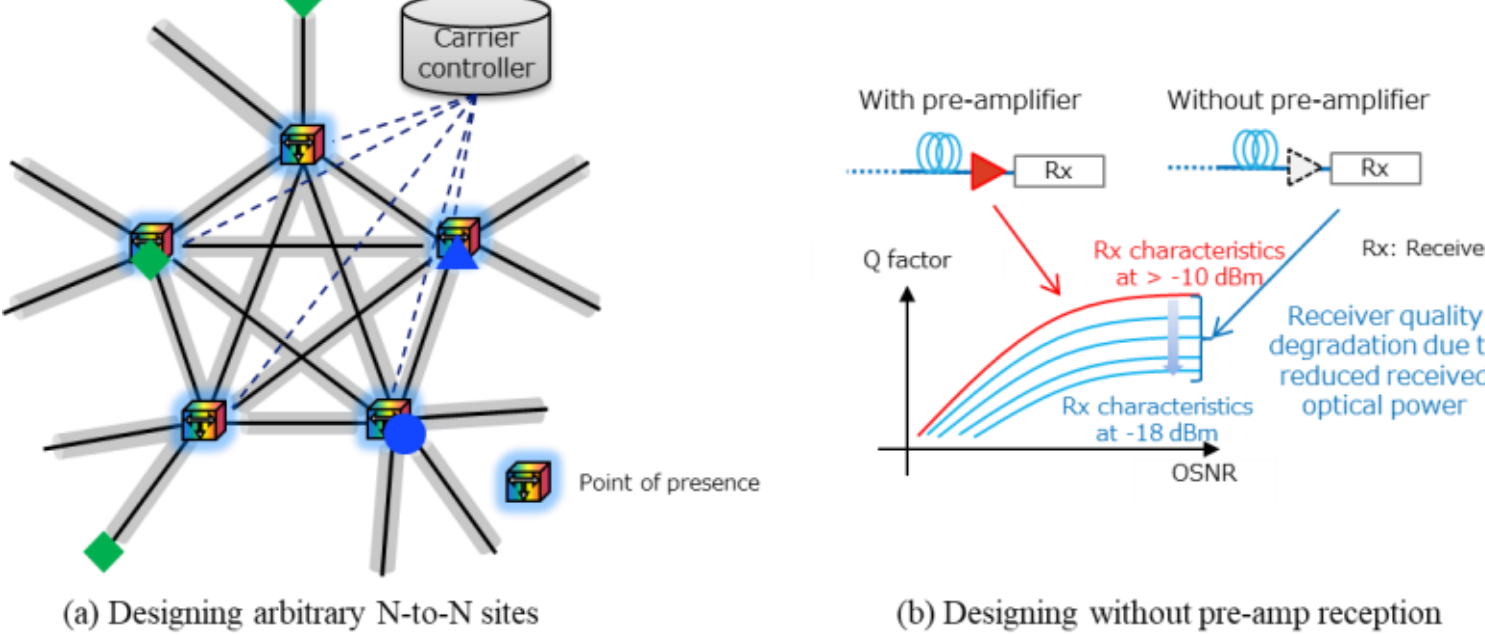

(a) Designing arbitrary N-to-N sites (b) Designing without pre-amp reception

Fig. 9. Design issues for DCX

Next, we focus on the Challenges 2) ~ 3). Fig. 9 presents the design issues associated with them. Fig. 9(a) illustrates the challenges involved in designing connectivity for arbitrary N-to-N sites. Compared with conventional P2P connection design, the complexity increases significantly; for example, when interconnecting five sites (green and blue) as shown in the figure, more than 40 possible design patterns exist. Fig. 9(b) shows an issue in which no pre-amplifier is installed at the receiver (Rx) and the Rx input power falls below around -10 dBm. In such cases, the design must explicitly account for QoT degradation caused by reduced Rx input power. It has been shown that signal quality degradation caused by TRxs can be modeled as additive and white Gaussian noise (AWGN), similar to amplified spontaneous emission (ASE) noise and nonlinear interference (NLI) noise [34]. Furthermore, degradation due to reduced received power can also be modeled as AWGN [35]. For this reason, the BER of the received signal can be evaluated by the aggregate amount of all impairment factors. Specifically, it can be expressed by the following equation:

$$BER = \kappa_1 erfc\left(\sqrt{\kappa_2 SNR}\right)$$

$\kappa_1$ and $\kappa_2$ are constants determined by the modulation format (e.g., in the case of 16QAM, $\kappa_1 = 3/8$, $\kappa_2 = 1/10$), and $erfc$ is the complementary error function. Here, the inverse of the overall system noise can be expressed as the sum of the inverses of the ASE noise term $SNR_{ASE}$, the NLI noise term $SNR_{NLI}$, the Rx-input-power-independent TRx noise term $SNR_{TRX'}$, and the Rx-input-power-dependent TRx noise term $SNR_P P_{in}$:

$$\begin{aligned} SNR^{-1} &= SNR_{ASE}^{-1} + SNR_{NLI}^{-1} + SNR_{TRX'}^{-1} + SNR_P^{-1} P_{in}^{-1} \\ &= GSNR^{-1} + SNR_{TRX'}^{-1} + SNR_P^{-1} P_{in}^{-1} \end{aligned}$$

This approach allows noise contributions from different physical impairments to be combined in a simple manner, with the resulting estimate matching the pre-FEC BER measured at the receiver and enabling practical use in design and maintenance operations. Next, we describe how this approach can be applied in practice to design. Fig. 10 presents an example of the design approach for arbitrary N-to-N sites shown in Fig. 9(a). Fig. 10(a) illustrates the possible patterns when two user sites (green) are interconnected via POPs. When the number of POPs between the user sites is limited to at most three to minimize latency, four patterns exist, as shown in Fig. 10(a): one using two POPs (blue line) and three using three POPs (red lines).

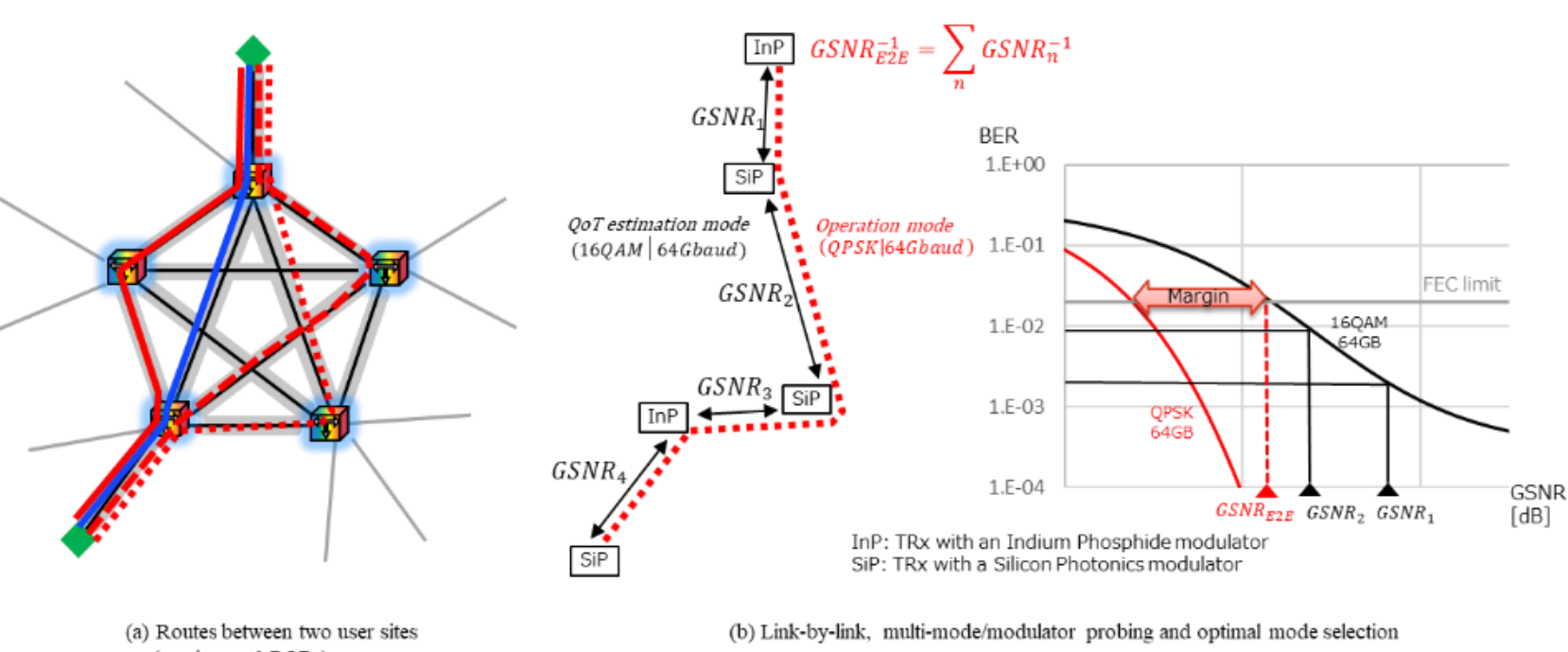

Fig. 10. An approach enabling fast and accurate selection of the optimal coherent mode.

The left side of Fig. 10(b) illustrates an example in which one of the routes shown by the red lines is divided into four segments, and QoT is estimated using TRxs equipped with different modulator technologies (Indium Phosphide, InP, and Silicon Photonics, SiP). The $GSNR_n$ of each segment can be obtained by subtracting the TRx noise from the $SNR_n$ measured on a per-segment basis using pre-FEC BER of 16QAM at 64 GBaud. By summing the inverses of the GSNRs of the individual segments, the inverse of the E2E GSNR can be derived. In this example, 16QAM at 64 GBaud is used during measurement, while QPSK at 64 GBaud is selected as the operational mode. [10] reports experimental results demonstrating the proposed design method using the National Science Foundation (NSF) COSMOS testbed in the United States, an urban-scale programmable testbed deployed on Manhattan Island in New York City. In this experiment, two types of CFP2-DCOs with different modulators (InP and SiP), six ROADMs, field fibers in a densely populated urban area, and fiber spools were used to construct routes of 72 km and 162 km. When the design method shown in Fig. 10 was applied for validation, the difference between the estimated and measured Q-factors was within 0.7 dB, even when the wavelength-dependent gain tilt of erbium-doped fiber amplifiers (EDFAs) was taken into account. Considering that the GSNR fluctuation of the experimental setup itself was on the order of 0.1~0.2 dB, this discrepancy can be regarded as sufficiently small. In this experiment, evaluations were also conducted using architecture and protocols that enable cooperation between users and carriers, as described in Fig. 8. These results indicate that the proposed approach can satisfy the requirements of Challenges 2)~3).

### *4.3 Line system*

DCX operators need to optimize optical wavelength paths across multiple line systems as described in the lower part of Fig. 3. The challenges and requirements can be summarized into the following three categories.

1) <u>Provision over access lines of unknown quality:</u> Ability to remotely monitor, optimize, and configure the line system without dispatching optical experts or adding extra hardware.
2) <u>Compatible with conventional operation methods while enabling fast provisioning and fault localization</u>: Engineers can visually confirm failures and roll back operations if unexpected issues occur.
3) <u>Responds to cases where multi-vendor line systems need to be constructed:</u> A physical model based approach allows estimation of all required link parameters, ensuring generalizability across equipment types and vendors.

The background for organizing the challenges into three categories is described below. Span loss measurements using optical time-domain reflectometry (OTDR) cannot be performed across EDFAs, so OTDR units must be deployed for each device, such as ROADMs or EDFAs, or field engineers must be dispatched to the locations where these devices are installed to perform measurements. Typically, ROADMs and EDFAs are autonomously controlled and optimized using vendor-specific Optical Supervisory Channels (OSCs), with each vendor employing its own proprietary methods. Although some aspects of OSC standardization—such as the wavelengths allocated for OSC channels—have been partially specified, comprehensive standardization has not been achieved. Consequently, it remains challenging to realize autonomous control and optimization across multi-vendor equipment. To support use cases illustrated at the bottom of Fig. 3, open and novel technologies and design methodologies that can be broadly applied to heterogeneous devices are required.

Beyond existing systems, novel approaches have been investigated in academic research. New optical path design methodologies based on the GN model have become widely adopted, leading to the development of open optical-networking tools such as GNPy [36] and Mininet-Optical [37]. By inputting system parameters of optical networks—such as fiber types and amplifier gains—into these tools based on the GN model, it is expected that the QoT can be analytically estimated with relatively low computational complexity [38]. In addition, ML has brought transformative advances across many scientific and engineering fields and achieved remarkable success. As a result, numerous studies on ML-based approaches for automated optimization of optical network operations have been reported over the past decade.

However, despite extensive research over many years, advanced automated operations enabled by ML have not yet been widely adopted in commercial optical fiber networks, and conventional line-system architectures continue to be deployed in practice. Based on the issues identified in [15, 21, 38], the challenges of ML-based autonomous operation of optical networks are analyzed from an operator's perspective as follows.

- **Transparency and explainability**: Many ML models contain a large number of parameters that interact in complex ways. As a result, it is often difficult to interpret how individual parameters or layers contribute to a given prediction, and the outcomes cannot always be explained in a simple cause-and-effect manner, leading to limited transparency and explainability.

- **Generalization and robustness**: Since ML models learn patterns from training data, their performance may degrade when the operational environment differs from the data distribution used during training, resulting in limited generalization capability.

DCX operators are required to guarantee network service quality in accordance with service-level agreements (SLAs) with users. In the event of failures, operators may be obligated to identify the root cause and the location of the failure and report them to users. However, black-box approaches are highly sensitive to temporal variations, link reconfigurations, changes in data characteristics, and differences in equipment types or vendors. Consequently, such approaches are not well-suited to meeting the requirements of DCX services, which must support heterogeneous line systems and enable on-demand service provisioning.

- **Cost constraints**: Many studies assume the deployment of numerous expensive measurement instruments, such as optical channel monitors (OCMs), within the network. However, the costs incurred by automation technologies—including newly installed equipment, expanded toolsets, and additional labor required for autonomous operation—must be justified by benefits that outweigh these expenditures.

Transition scenarios from existing operational practices are also an essential factor. For example, in the context of autonomous driving, levels 1 through 5 are defined according to the degree of automation. Such staged deployment is essential for ensuring safety, regulatory compliance, and societal acceptance during technology adoption. Similar stepwise transition processes are required for DCX services, which are expected to support critical social infrastructure.

- **Compatibility with existing operational practices**: Optical transmission engineers are highly experienced with well-established conventional, non-automated operational methods that are routinely applied in practice. To introduce new technologies and operational paradigms, clear migration scenarios from existing operations must be provided.

Based on these challenges and requirements, we set the following targets. All physical parameters required for optical path design are extracted from the line system and visualized, enabling supervisors located in different time zones to perform design and analysis using open tools and to decide whether to complete or roll back the deployment [15]. The additional hardware required for automation and optimization is limited to approximately one set per line system, and this hardware can be reused for the optimization of other line systems. The time required to optimize already deployed equipment is constrained to within one hour, corresponding to less than half of a typical two-hour maintenance window. In terms of accuracy, the target is to achieve a design error no greater than that obtained when optimization is performed through on-site engineering work.

To address Challenges 1)~3) and meet the targets described above, we have been investigating integrated physical-aware methods in Fig. 11 that combine multiple monitoring technologies. These methods extract link parameters and perform automated optimization based on physical models, and are built upon two core techniques: DLM and OLS calibration.

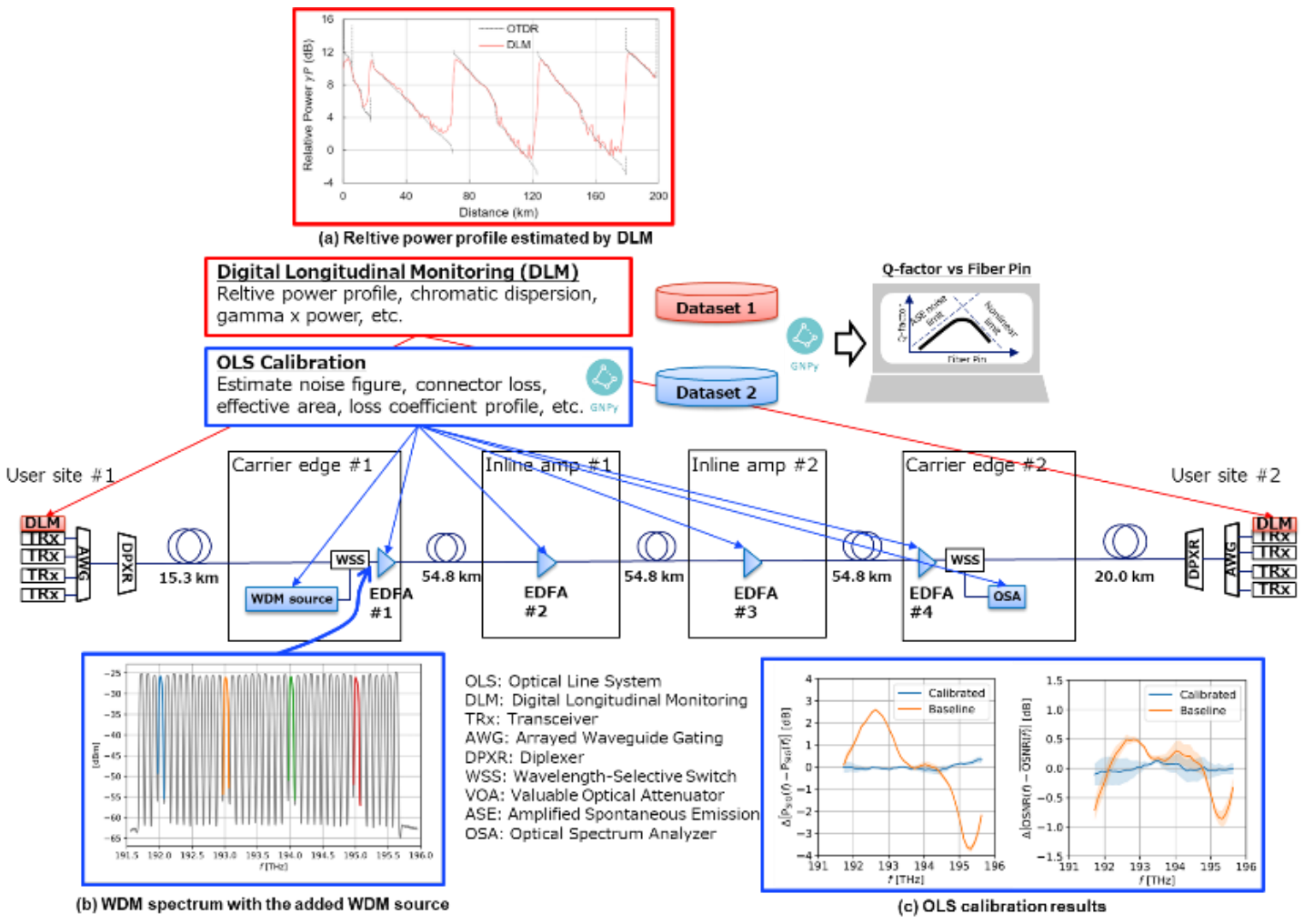

Fig. 11. Integrated physical-aware method. DLM functions are implemented in the TRxs at user site #1 and #2 to estimate parameters such as relative power profiles and chromatic dispersion. (a) shows the relative power profile estimated by DLM (red line) with the OTDR measurement result (black dashed line). At carrier edge #1 and #2, a WDM source and an optical spectrum analyzer (OSA) are deployed, respectively. The OLS calibration function controls these devices together with EDFA #1~#4, and applies feedback with GNPy to estimate parameters such as noise figure and connector loss, thereby optimizing the gain and tilt of the EDFAs. (b) shows the WDM spectrum with the added WDM source (gray), while (c) compares the effects of OLS calibration (signal power Pin and OSNR) against the baseline [15].

*Digital Longitudinal Monitoring (DLM):* DLM is a transponder-based monitoring technique that visualizes physical characteristics of multiple link components along the fiber-longitudinal direction solely by processing received data-carrying signals [11]. The red parts of Fig. 11 illustrate the relative power profile estimated by DLM. With a priori information of fiber-span lengths, span-wise CD parameters can also be estimated by calibrating positions of EDFAs, indicated by longitudinal power profiles. Unlike OTDR that requires a dedicated measuring device for each EDFA, DLM supports multi-span measurements, enabling rapid and cost-efficient monitoring of line systems. Furthermore, by providing E2E visibility of the entire line system, DLM facilitates the optimization and localization of impairments and failures originating from line system components, including fibers, amplifiers, and filters. In addition, the distance-resolved power profile can be visualized in multiple dimensions—including frequency, polarization, and time—by employing multiple channels, polarization-resolved measurements, or continuous temporal visualization, a concept referred to as optical link tomography (OLT) [12]. Furthermore, in [13], the network-wide optical tomography across multiple domains, including a live production network, has been demonstrated. By visualizing E2E optical powers with DLM/OLT along multiple routes across multiple domains with commercial transponders, performance bottleneck localization, power and routing optimization were successfully achieved, allowing for connectivity of a 400G real-time TRx with an increased margin.

*Optical Line System (OLS) calibration*: OLS calibration (blue parts of Fig. 11) is a telemetry-based technique improving QoT estimation by determining the physical parameters using only optical power monitors integrated in EDFAs and one optical spectrum analyzer (OSA) at both edge nodes of the considered optical line [39]. The retrieved physical parameters can then be applied in a physics simulation model, such as GNPy, to define the optimal operation working point.

The physical parameters extracted by these two techniques are stored in Datasets 1 and 2 respectively, and by feeding this information into GNPy, the graphs required for optical path design such as power profile vs distance and Q-factor vs fiber-input-power can be visualized [15].

The tasks of parameter retrieval with DLM and OLS calibration, equipment setting optimization, human check/decision using the visualized graphs, and system setup/provisioning were completed within 1 hour, with field fiber network facilities at Duke University, Durham, North Carolina [15]. By comparing Q-factor estimates calculated from the extracted link parameters with measured results from 400G TRxs, we confirmed that our methodology has a reduction in the QoT prediction errors ($\pm$ 0.3 dB) over existing design commonly used in on-site design work ($\pm$ 0.6 dB).

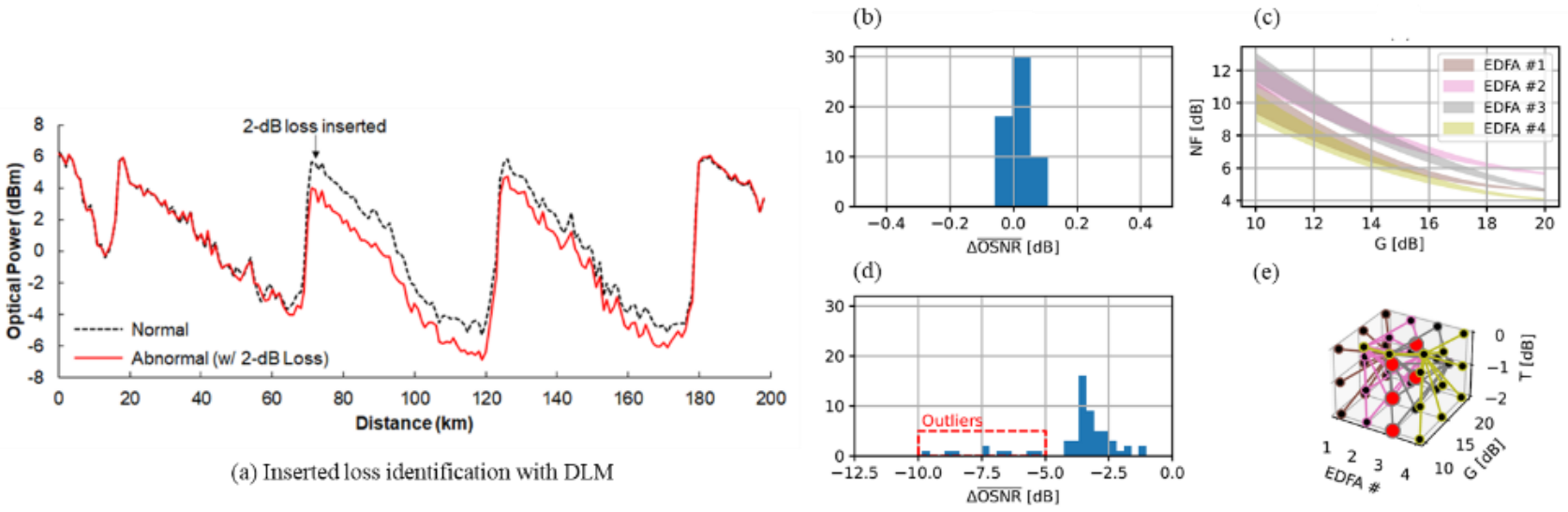

Fig. 12. Fault localization with the integrated physical-aware method. (a) shows the output power result in which a 2 dB loss was inserted after EDFA #2. (b) ~ (e) shows EDFA #3 noise figure degradation fault: (b) Average OSNR error distribution before fault; (c) average noise figure function vs. gain target before fault; (d) average OSNR error distribution after fault; (e) fault localization throughout outliers identification.

In Challenge 2), the requirement for detecting and localizing a fault is also emphasized. This paragraph discusses the application of fault localization by leveraging the integrated physical-aware method to reveal changes in physical metrics and parameters. Using the experimental setup shown in Fig. 11, DLM was performed with intentionally introduced loss. The black dotted line in Fig. 12(a) represents the condition before loss insertion, while the red line shows the DLM result after a 2 dB loss was inserted immediately after EDFA #2. The results clearly demonstrate that both the magnitude and the location of the loss inserted can be accurately identified. Next, we discuss the detection of noise figure degradation. Fig. 12(b) and (c) show the average OSNR error distribution $\Delta\overline{OSNR}$, and the noise figure vs. gain of each EDFA between the measured one and the GNPy prediction with the physical parameters derived through the OLS calibration methodology under normal conditions before the failure. We inserted 8 dB attenuator into the intermediate stage of the EDFA #3 to modify the noise figure degradation. After failure, the same dataset is collected and distribution of the average OSNR error is evaluated again (d): the distribution shows the presence of outliers along the tails that were not present before [40]. Therefore, it is possible to localize the EDFA that suffered the fault, as the OLS calibration procedure is sensitive to the change of the noise figure function in shape (e).

In the previous paragraph, we discussed the technologies and their validation results for Challenges 1) and 2). In this paragraph, we present the latest initiatives and validation results related to Challenge 3).

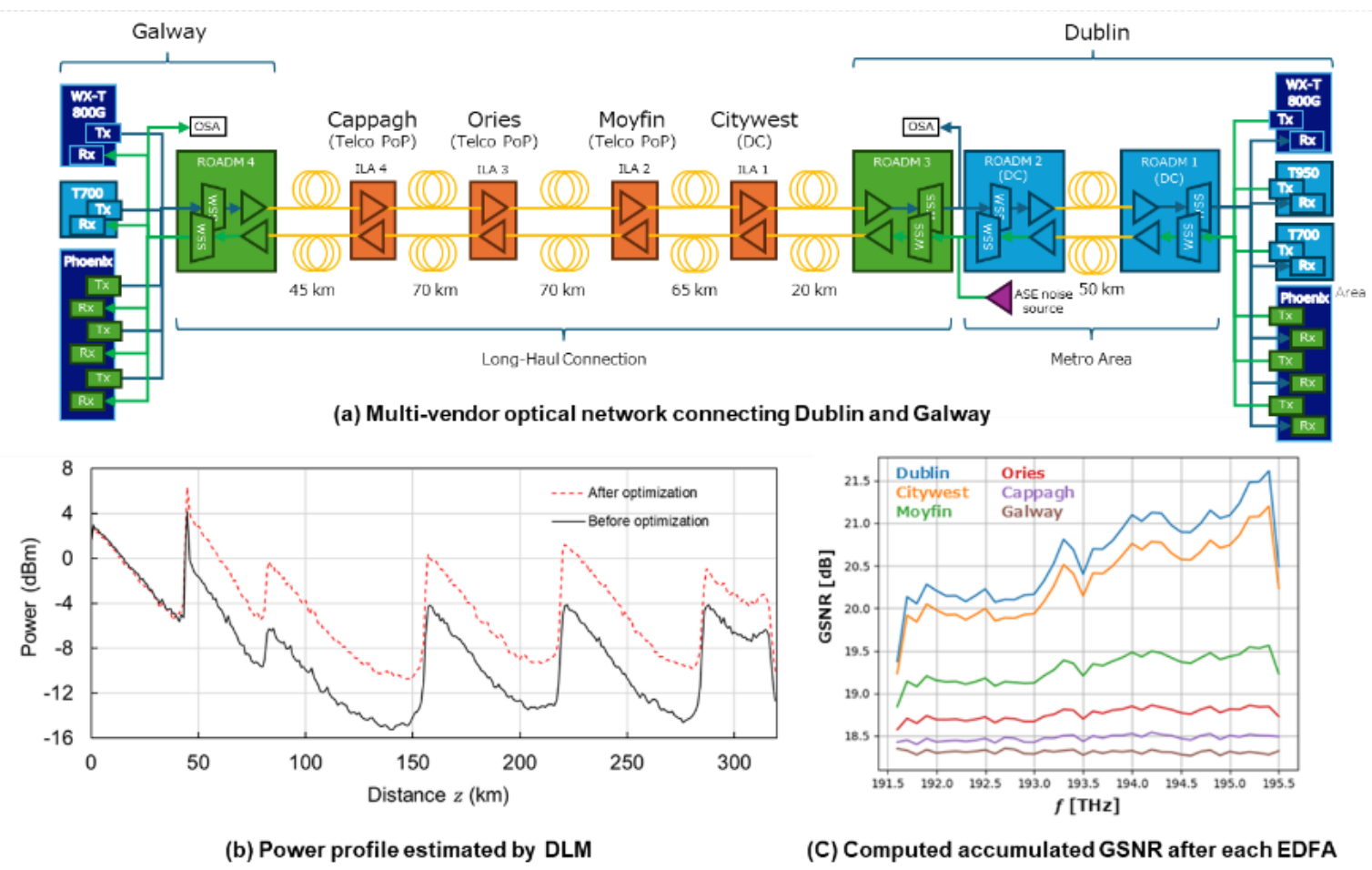

Fig. 13. Line-system optimization experiment. (a) experimental network configuration from Galway to Dublin with multi-vendor devices including four common in-line amplifiers. (b) power profiles estimated by DLM, (c) accumulated GSNR after each EDFA in (a).

Fig. 13 illustrates the experimental results demonstrating fast optimization of a multi-vendor line system [14]. Fig. 13(a) shows an experimental network spanning approximately 280 km between Dublin and Galway. The color variations of the devices indicate vendor difference; the long-haul network was built using the ROADM greyboxes [41] and common EDFAs, extending from the Dublin metro network to Galway. The goal was to rapidly optimize a temporary WDM link during a large-scale disaster using dark fiber and borrowed multi-vendor devices. After installing four in-line amplifiers (ILAs) in their initial configuration, the DLM measurement before optimization is shown as the black line in Fig. 13(b), while the DLM measurement after optimizing the EDFAs using OLS calibration is shown as the red dashed line. Fig. 13(c) illustrates the accumulated GSNR at the output of each EDFA, optimized so that wavelength dependency becomes flat at the final EDFA in Galway. No special EDFAs or OSC were used; only the total input/output power of the EDFAs was monitored from the Dublin site, and adjustments were limited to Gain/Tilt setting.

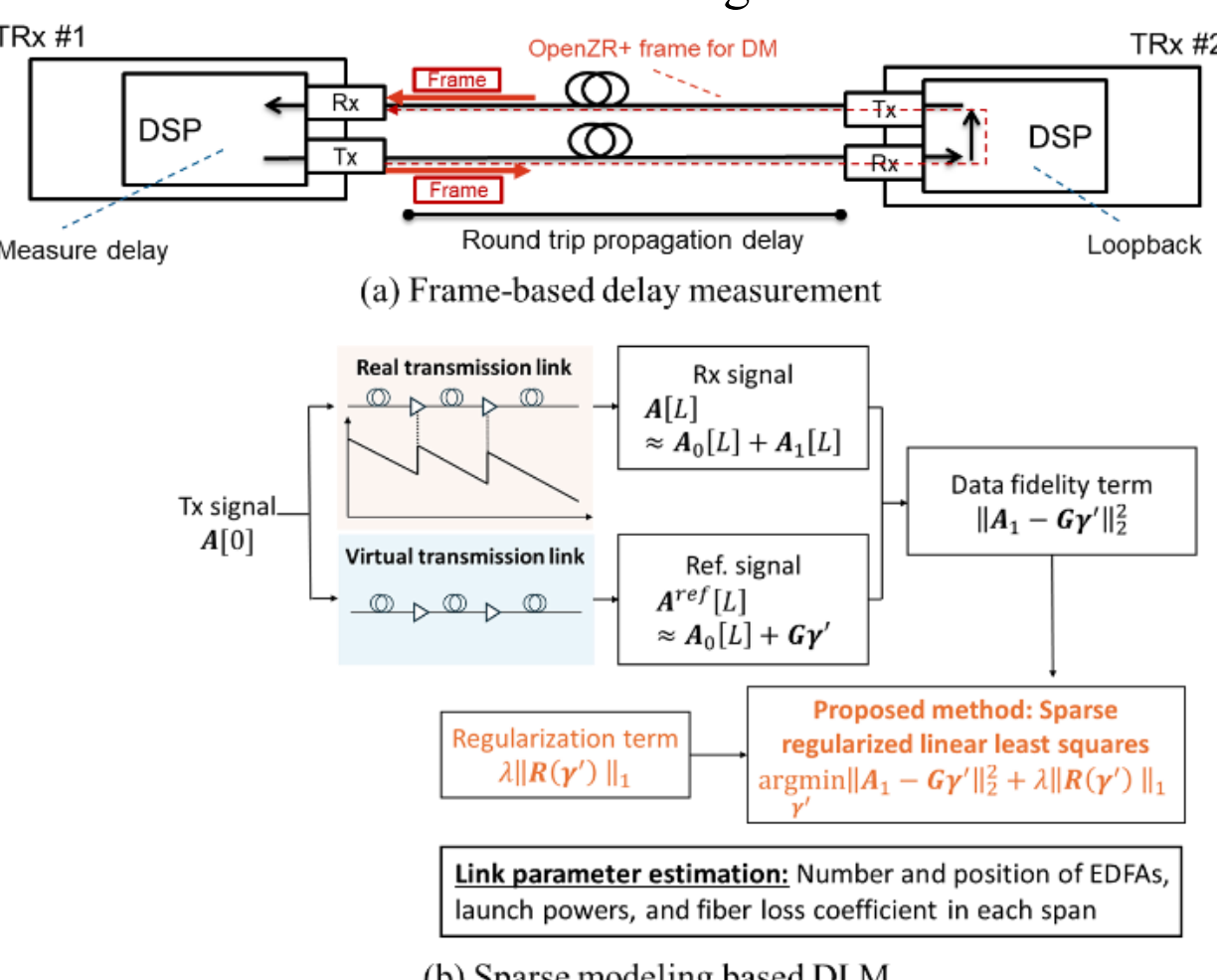

Fig. 14. TRx-based physical-parameter estimation technologies.

Fig. 14 shows the new TRx-based physical-parameter estimation technologies, the frame-based delay measurement and the sparse modeling based DLM. When estimating the total span length, delay measurement based on OTN frames is commonly used in telecom networks. However, this approach cannot be applied in DCX when Ethernet-based IPoDWDM equipment using QSFP-DD modules is employed. Therefore, a delay measurement method based on the CMIS is required. Fig. 14(a) shows frame-based delay measurement using an Open ZR+ pluggable TRx. We inserted a dedicated measurement bit sequence into the reserved overhead of the Open ZR+ frame, transmitted it in a loopback configuration by returning it at the remote endpoint TRx, and measured the round-trip delay as the elapsed time until the returned measurement sequence was detected [42]. Using this approach, total span-length estimation becomes feasible even when Ethernet-based IPoDWDM devices are involved.

We also introduce a new technical approach related to physical-parameter estimation using DLM. To accurately estimate physical-parameters, it is essential to extract low-noise power profile. Conventionally, low-noise power profile was estimated by averaging multiple power profiles; however, this approach poses a challenge in terms of long measurement times. To address this issue, Fig. 14(b) presents a robust DLM approach using sparse modeling that improves estimation accuracy with few measurements. This method integrates noise reduction of power profile and physical-parameter estimation in a mutually reinforcing loop as sparse regularization: preliminary noise removal allows for an initial physical-parameter estimate, which then acts as a physical constraint to further refine the power profile and ultimately finalize the physical-parameter estimation. In [43], results show that this approach improves power-profile estimation accuracy by 0.4 dB and successfully detects anomalous losses as small as 0.72 dB using two-stage sparse regularization. Furthermore, this approach works with few measurements, reducing the DLM computation time to 1/174 of that required by conventional power profile averaging. In the context of ONDT operation, the more significant advantages demonstrated in the same work include compression of DLM computation and the ability to estimate the number and position of EDFAs, launch powers, and fiber loss coefficient in each span.

As demonstrated in this section, TRx-based physical-parameter estimation technologies incorporating DLM and OLS calibration enable automated monitoring/estimation and optimization of line-system parameters in a vendor-independent manner, facilitating openness and interoperability across heterogeneous systems.

## Future research

In the context of a user–carrier collaborative approach, promising research directions include protocol development for rerouting in the event of failures, OpenConfig-based control in multi-vendor environments, and autonomous control leveraging DLM and other physical-parameter extraction techniques. Continuous efforts toward standardization are essential for multi-vendor ROADM control and operation of it. Regarding the utilization of ML, recent studies [44] have advanced techniques that apply hybrid analytical/ML models for OSNR estimation, and determining the domains in which ML should be effectively applied remains an intriguing research challenge.

## Conclusion

We have proposed the DCX as one of the approaches to support All-Photonics Networks-as-a-Service, which enables a virtual large-scale DC by directly interconnecting multiple geographically distributed DCs in metropolitan areas. DCX can flexibly and rapidly establish optical networks by interconnecting devices from multiple vendors between any multi-point locations, using various transmission modes suited to link distances and QoT. In this paper, we first examine the key drivers behind considering DCX, focusing on technological advancements and changes in the surrounding environment. We then introduce the proposed DCX architecture and clarified its distinguishing characteristics with respect to existing optical networking approaches, including its operator-driven and programmable design, user–carrier collaborative control framework, and support for alien access links with unknown physical parameters. Based on this comparison, we identify the following key requirements for realizing

DCX: supporting low-latency operations, scalability, reliability, and flexibility within a single network architecture; enabling the addition of new operator-driven automation functionalities; and enabling control and management of remotely deployed transponders connected via access links whose parameters are unknown. Next, we analyze the technical challenges for fulfilling these requirements from the perspectives of optical transponders, optical controllers, and OLSs. As solutions to these challenges, we propose a set of technologies that enable ONDT operations, including cloud-native architecture for coherent TRx, remote transponder control, fast E2E optical path provisioning, TRx-based physical-parameter estimation technologies incorporating DLM, and OLS calibration. Finally, we demonstrate the feasibility of integrating these technologies into a unified, operator-driven and open digital twin framework that spans both transponders and line systems. This cross-domain and system-level integration has the potential to enable scalable, practical, and vendor-agnostic operation in heterogeneous and multi-operator optical networks, and represents a key step toward realizing DCX in real-world deployments.

### Funding

These research results were obtained through work founded by the grant program (No. JPJ012368G50201) of the National Institute of Information and Communications Technology, Japan, the National Science Foundation (CNS-1827923, CNS-2112562, CNS-2128638, CNS-2148128, CNS-2211944, CNS-2330333, EEC-2133516, OAC-2029295), and Taighde Éireann: 13/RC/2077_P2, 18/RI/5721, 22/FFP-A/10598.

### Acknowledgements

The authors would like to thank Masahisa Kawashima, Soichiroh Usui, and Masatoshi Namiki of NTT; Vittorio Curri of Politecnico di Torino; and Daniel Kilper and Marco Ruffini of Trinity College Dublin for their valuable discussions that contributed to this work. This research was conducted in cooperation with Politecnico di Torino, Columbia University, Duke University, and Trinity College Dublin. We thank Rob Lane and the CRF team (Columbia), Lumentum, FOC, the Telecom Infra Project OOPT-PSE working group, and the Duke Office of Information Technology (OIT) team, HEAnet for their support.